\documentclass[12pt, reqno]{article}
\usepackage{epsfig}
\usepackage{amssymb}
\usepackage{amsmath}
\usepackage{mathrsfs}
\usepackage{todonotes}
\usepackage{slashed}
\usepackage{amsmath}	
\usepackage{amsthm}		
\usepackage{amssymb}	
\usepackage{mathtools}	
\usepackage{enumerate}	
\usepackage{paralist}		
\usepackage[a4paper,top=1in,bottom=1in,left=1in,right=1in]{geometry}	
\usepackage{xcolor}
\usepackage{bbold}
\usepackage{bigints}

 \usepackage[unicode=true,
              colorlinks=true,
              linkcolor=blue,
              citecolor=red]{hyperref}

\newcommand{\be}{\begin{eqnarray}}
\newcommand{\ee}{\end{eqnarray}}

\def\cs{{\cal S}}

\begin{document}

\baselineskip=18pt

\setcounter{footnote}{0}
\setcounter{figure}{0}
\setcounter{table}{0}

\begin{titlepage}
\unitlength = 1mm
\ \\

\today
\begin{center}

{ \LARGE {\textsc{Soft Graviton Theorem in Arbitrary Dimensions}}}

\vspace{0.8cm}
Nima Afkhami-Jeddi$^*$

\vspace{1cm}
\begin{abstract}
In this note we show that the recent conjecture proposed by Cachazo and Strominger holds at tree level in arbitrary dimensions. The proof makes crucial use of the fact that the sub-leading operator is defined using the total angular momentum operator. A key ingredient that makes the proof possible is the CHY formula for graviton amplitudes in arbitrary number of dimensions. 
\end{abstract}
\vspace{0.5cm}

\vspace{1.0cm}

\end{center}
\vspace{6.0cm}
$^*${\it Perimeter Institute for Theoretical Physics, Waterloo, ON, Canada} \\

\end{titlepage}

\pagestyle{empty}
\pagestyle{plain}

\def\ip{${\cal I}^+$}
\def\p{\partial}
\def\cs{{\cal S}}

\def\co{{\cal O}}

\pagenumbering{arabic}

\tableofcontents

\section{Introduction}
Weinberg showed in the 60's that a graviton scattering amplitude has a universal factorization behaviour in the so called soft limit of taking the n-th particle momentum to be near zero \cite{weinberg}. That is as $k_n\rightarrow0$:
\begin{align}
\mathcal{M}_n(k_1,k_2,...,k_n)=\sum\limits_{a=1}^{n-1}\frac{E_{\mu\nu}k_a^\mu k_a^\nu}{k_n.k_a}\mathcal{M}_{n-1}(k_1,k_2,...,k_{n-1})+\mathcal{O}(k_n),
\end{align}
where $E_{\mu\nu}$ is the polarization tensor for the n-th graviton and $\mathcal{M}_n$ denotes the complete n-particle graviton amplitude including the momentum conserving delta distribution. Recently Cachazo and Strominger provided evidence for a conjecture \cite{cachstro} stating that the terms of higher order in this expansion also have a universal behaviour. Furthermore, using the BCFW relations and spinor helicity formalism they showed that the conjecture holds non-trivially for the case of tree level amplitudes in four dimensions. The conjecture may be denoted in the following way:
\begin{align}
\label{conjec1}
\mathcal{M}_n(k_1,k_2,...,\lambda k_n)=\left(\frac{1}{\lambda}S^{(0)}+S^{(1)}+\mathcal{O}(\lambda)\right)\mathcal{M}_{n-1}(k_1,k_2,...,k_{n-1}).
\end{align}
The first term on the right hand side is given by the well known Weinberg soft limit formula and the subleading terms are given by the action of operators that are constructed from the total angular momentum operator whose explicit form is given in subsequent sections. As will be shown in section \ref{secexp}, the calculation of the right hand side naturally separates into two parts. One part involves computing the action of the orbital angular momentum operator on delta functions of the so called scattering equations. The second part involves computing the action of the total angular momentum operator including both the orbital and spin contributions on the determinant of a matrix composed of Lorentz contractions between momenta and polarization vectors of external particles.

In this paper we will use the Cachazo, He, and Yuan formula for the d-dimensional tree level graviton amplitude \cite{chyklt} to show that the conjecture holds non-trivially in higher dimensions. The proof proceeds by explicitly computing the $\lambda$ expansion on the left hand side of \eqref{conjec1} and comparing the resulting expression with the explicit calculation of the right hand side. 

This paper is organized as follows. In section \ref{secchy} we review the CHY formula for the tree level graviton scattering amplitude. In section \ref{secexp} we will perform the explicit expansion on the left hand side of equation \eqref{conjec1} and show that the Weinberg soft limit term as well as higher order corrections may be obtained by a contour deformation and Cauchy's residue theorem. In section \ref{secs1ondet} we will explicitly compute the higher order terms on the right hand side of equation \eqref{conjec1} and show that the two expressions agree and thereby complete the proof.

{\bf Note added:} As already mentioned, completing the proof involves computing the action of the total angular momentum operator including the spin and orbital contribution. While this work was being prepared for submission, a paper \cite{schwab} appeared in which part of the soft theorem corresponding to the orbital angular operator was proven using similar methods to those used here.
\section{Tree level graviton scattering and CHY formula}
\label{secchy}
A set of equations relating the position of n punctures on a sphere to the space of kinematic invariants of an n-particle massless scattering amplitude was found by Cachazo, He, and Yuan(CHY)\cite{chyklt}. These equations are called the scattering equations and are given by:
\begin{align}
&\sum\limits_{b\neq a}\frac{k_a\cdot k_b}{\sigma_a-\sigma_b}=0& \forall a\in \{1,2,...,n\}.
\end{align}
In this section and the rest of the paper we will adopt the notation used in \cite{chyexp}. Using these equations the existence of a closed form equation for the complete tree level S-matrix of gravity theories in any dimension was proposed by CHY \cite{chyklt} as an integration over the variables $\sigma_i$ localized on the solutions of the scattering equations. The explicit construction was subsequently outlined in \cite{chyexp}. In this section we will briefly review this construction in which a n-particle graviton scattering amplitude is given by the following equation:
\begin{align}
M_{n}=&\bigintss \frac{d^n\sigma}{vol SL(2,\mathbb{C})}\left[ \prod\limits_{a}{'}\delta\left(\sum\limits_{b\neq a}\frac{k_a.k_b}{\sigma_{a}-\sigma_{b}}\right)\right]E_n(\{k\},\{\epsilon\},\{\sigma\}).
\end{align}
Since the scattering equations are invariant under $SL(2,\mathbb{C})$ transformationsz only $n-3$ of these equations are independent. Using a Fadeev-Popov procedure to mod out this redundancy the amplitude may be written as:
\begin{align}
M_{n}=&\bigintss \left[ \prod\limits_{c\neq p,q,r}d\sigma_c\right](\sigma_{pq}\sigma_{qr}\sigma_{rp})(\sigma_{ij}\sigma_{jk}\sigma_{ki}) \left[ \prod\limits_{a\neq i,j,k,n}\delta\left(\sum\limits_{b\neq a}\frac{k_a \cdot k_b}{\sigma_{ab}} \right) \right] E_n(\{k\},\{\epsilon_i\},\{\sigma\}),
\end{align}
where $\sigma_{ab}\equiv \sigma_a-\sigma_b$ and $E_n$ is a function of the integration variables $\sigma_i$ as well as the momentum and the polarization vectors of the external particles involved in the scattering process. It is clear that the delta functions completely localize all the integrals and the integration reduces to a sum of the integrand as well as a Jacobian factor on solutions of the scattering equations.
To give the explicit form of $E_n$ ,we introduce the following definition:
\begin{align}
\Psi\equiv 
\left(
\begin{array}{cc}
A & -C^T \\ 
C & B
\end{array} 
\right),
\end{align}
where $A,B,C$ and $D$ are $n\times n$ matrices whose components are given by the following:
\begin{align}
&A_{ab}=\left\{ \begin{array}{cc}
\frac{k_a.k_b}{\sigma_{ab}}&a\neq b \\ 
0&a=b
\end{array}, \right.
&B_{ab}=\left\{ \begin{array}{cc}
\frac{\epsilon_a.\epsilon_b}{\sigma_{ab}}&a\neq b \\ 
0&a=b
\end{array}, \right.\hspace{.3in}
&C_{ab}=\left\{ \begin{array}{cc}
\frac{\epsilon_a.k_b}{\sigma_{ab}}&a\neq b \\ 
\sum\limits_{c\neq a}\frac{\epsilon_a.k_c}{\sigma_{ca}}&a=b
\end{array}. \right.
\end{align}
With this definition $E_n$ is given by the following expression :
\begin{align}
\label{deteqn}
E_n\equiv \frac{4}{(\sigma_{i}-\sigma_j)^2}\det{(\Psi^{ij}_{ij})},
\end{align}
Where $\Psi^{ij}_{ij}$ is the matrix obtained by removing the rows $i,j$ and the columns $i,j$ from the matrix $\Psi$. It was shown in \cite{chyexp} that the result is independent of the choice of $i$ and $j$.
\section{Taking the soft limits and expansion in $\lambda$}
\label{secexp}
\indent Weinberg's soft limit theorem \cite{weinberg} states that the single soft limit of a graviton scattering amplitude to lowest order is given by the universal Weinberg pole formula. The correct soft limit behaviour of CHY equations to leading order was demonstrated in \cite{chyklt,chyexp}. More recently Cachazo and Strominger presented evidence for a conjecture giving a universal formula for higher order terms in the soft limit of a graviton scattering amplitude \cite{cachstro}. In this paper they showed using the BCFW recursion relations that this conjecture holds non-trivially for all tree level graviton scattering amplitudes in four dimensions. The conjecture can be stated as follows:
\begin{align}
\label{conj}
\mathcal{M}_{n+1}(k_1,k_2,...,k_n,\lambda k_{n+1})=\left(\frac{1}{\lambda}S^{(0)}+S^{(1)}\right)\mathcal{M}_n(k_1,k_2,...,k_n)+\mathcal{O}(\lambda),
\end{align}
where $\lambda$ is the expansion parameter and $\mathcal{M}_{n+1}$ is the complete expression for the scattering amplitude including the momentum conserving Dirac-delta distribution. The first term on the right hand side of \eqref{conj} is given by Weinberg's soft theorem and $S^{(1)}$ in the second term is an operator acting on the scattering amplitude. The explicit form of $S^{(1)}$ is given by:
\begin{align}
S^{(1)}&=\sum\limits_{a=1}^n\frac{E_{\alpha\nu}k_a^\alpha k_{n,\mu} \hat{J}_a^{\mu\nu}}{k_n.k_a},
\end{align}
where $E_{\mu\nu}$ is the polarization vector of the soft graviton and $\hat{J}_a^{\mu\nu}$ is the complete angular momentum operator, including both the orbital and spin contributions. 

In order to show that the conjecture holds in all dimensions at tree level we will use the CHY formula for the graviton scattering amplitude. By computing the expansion of the scattering amplitude in the soft limit parameter $\lambda$ and comparing the results with the action of $S^{(1)}$ on the scattering amplitude on the right hand side of equation \eqref{conjec1} we will confirm the conjecture for tree level processes.

To make the expressions more compact we make the following definitions :
\begin{align}
 f_a^n&\equiv \sum\limits_{b\neq a}^n\frac{k_a.k_b}{\sigma_{ab}},\notag\\
 \bigintss D\sigma&\equiv\bigintss \left[ \prod\limits_{c\neq p,q,r,n}d\sigma_c\right](\sigma_{pq}\sigma_{qr}\sigma_{rp})(\sigma_{ij}\sigma_{jk}\sigma_{ki}).
\end{align}

First consider an n-particle graviton scattering amplitude with the soft limit taken on the n-th particle $(k_n\rightarrow\lambda k_n)$:
\begin{align}
M_{n}=&\bigintss \left[ \prod\limits_{c\neq p,q,r,n}d\sigma_c\right](\sigma_{pq}\sigma_{qr}\sigma_{rp})(\sigma_{ij}\sigma_{jk}\sigma_{ki})\left[ \prod\limits_{a\neq i,j,k,n}\delta\left(\sum\limits_{b\neq a,n}\frac{k_a.k_b}{\sigma_{ab}}+\frac{\lambda k_a.k_n}{\sigma_{an}}\right)\right]\notag&\\
&\hspace{2cm}\times \bigintss d\sigma_n \frac{1}{\lambda}\delta \left(\sum\limits_{b\neq n}\frac{k_n.k_b}{\sigma_{nb}}\right)E_n(\{k\},\{\epsilon_i\},\{\sigma\},\lambda).
\end{align}
The expansion of the delta functions corresponding to the first $n-1$ scattering equations is given by
\begin{align}
\prod\limits_{a\neq i,j,k,n}&\delta\left(
\sum\limits_{b\neq a,n}\frac{k_a.k_b}{\sigma_{ab}}+\frac{\lambda k_a.k_n}{\sigma_{an}}\right)
=\prod\limits_{a\neq i,j,k,n}\left[
\delta
\left(
\sum\limits_{b\neq a}^{n-1}\frac{k_a.k_b}{\sigma_{ab}}
\right)+
\lambda \frac{k_a.k_n}{\sigma_{an}}\delta{'}
\left( 
\sum\limits_{b \neq a}\frac{k_a.k_b}{\sigma_{ab}}\right)+\mathcal{O}(\lambda^2)
\right]\notag&\\
=\prod\limits_{a\neq i,j,k,n}&\delta
\left(
f_a^{n-1}
\right)+\lambda \sum\limits_{l\neq i,j,k,n}\left[ \prod\limits_{a\neq i,j,k,n,l} \delta\left(
f_a^{n-1}\right)
\right]
\frac{k_n.k_l}{\sigma_{ln}}\delta{'}\left(
f_l^{n-1}
\right)+\mathcal{O}(\lambda^2)\notag&\\ 
\notag&\\
\equiv\delta^{(0)}+&\lambda\delta^{(1)}+\mathcal{O}(\lambda^2).
\end{align}
Noting the definitions made in the last line and dropping terms of $\mathcal{O}(\lambda)$ ,we may rewrite the expansion of the scattering amplitude as:
\begin{align}
\label{exp1}
M_{n}&=\bigintss D\sigma\left[
\frac{\delta^{(0)}}{\lambda}+\delta^{(1)}
 \right]
\bigintss d\sigma_n \delta \left(\sum\limits_{b\neq n}\frac{k_n.k_b}{\sigma_{nb}}\right)\left(E_n^{(0)}+\lambda E_n^{(1)}\right)\notag&\\
&=\frac{1}{\lambda}\bigintss D\sigma\delta^{(0)}\bigintss d\sigma_n \delta \left(\sum\limits_{b\neq n}\frac{k_n.k_b}{\sigma_{nb}}\right)E_n^{(0)}\notag&\\
&+\bigintss D\sigma\delta^{(1)}\bigintss d\sigma_n\delta \left(\sum\limits_{b\neq n}\frac{k_n.k_b}{\sigma_{nb}}\right)E_n^{(0)}\notag&\\
&+\bigintss D\sigma\delta^{(0)}\bigintss d\sigma_n\delta \left(\sum\limits_{b\neq n}\frac{k_n.k_b}{\sigma_{nb}}\right)E_n^{(1)},
\end{align}
Where $E_n^{(i)}$ is the i-th coefficient in the Taylor expansion of $E_n$ in $\lambda$. Note that the first term is given by Weinberg's soft theorem:
\begin{align}
 \stackrel{\mbox{$M_n$}}{\mbox{\tiny$\lambda\rightarrow0$}}
&=\frac{1}{\lambda}\bigintss D\sigma\delta^{(0)}\bigintss d\sigma_n \delta \left(\sum
\limits_{b\neq n}\frac{k_n.k_b}{\sigma_{nb}}\right)E_n^{(0)}\notag&\\
&=\frac{1}{\lambda}\bigintss D\sigma\delta^{(0)}\bigintss d\sigma_n \delta \left(\sum\limits_{b\neq n}\frac{k_n.k_b}{\sigma_{nb}}\right)C_{nn}^2 E_{n-1}\notag&\\
&=\frac{1}{\lambda}\sum\limits_{b\neq n}\frac{(\epsilon_n.k_b)^2}{k_n.k_b}M_{n-1}.
\end{align}
The full expression for the second term of \eqref{exp1} is given by:
\begin{align}
\label{secondterm}
&\bigintss D\sigma\delta^{(1)}\bigintss d\sigma_n\delta \left(\sum\limits_{b\neq n}\frac{k_n.k_b}{\sigma_{nb}}\right)E_n^{(0)}\notag&\\
&=\bigintss \left[ \prod\limits_{c}{'}d\sigma_c\right]\sum\limits_{l\neq i,j,k,n}\left[ \prod\limits_{a\neq i,j,k,n,l} \delta\left(
f_a^{n-1}\right)
\right]
\frac{k_n.k_l}{\sigma_{ln}}\delta{'}\left(
f_l^{n-1}
\right)E_n^{(0)}.
\end{align}
Note that the integral over $\sigma_n$ does not have a pole at $\infty$ and therefore we may use contour deformation to evaluate this integral using a residue theorem.
To prove the conjecture we must consider the action of the operator $S^{(1)}$ on the scattering amplitude $M_{n-1}$. In order to do this we write out the operator $S^{(1)}$ as it acts on the scattering equations:
\begin{align}
S^{(1)}&\equiv\sum\limits_{a=1}^{n-1}\frac{E_{\mu\nu}k_a^\mu k_{n\rho} \hat{J}_a^{\rho\nu}}{k_n.k_a}
=\sum\limits_{a=1}^{n-1}\left((\epsilon_n.k_a) \epsilon_\nu\frac{\partial}{\partial k_{a\nu}}-\frac{(\epsilon_n.k_a)^2}{k_n.k_a}k_{n\rho}\frac{\partial}{\partial k_{a\rho}}\right).
\end{align}
The action of the operator on the scattering equations is then given by:
\begin{align}
S^{(1)}f_l^{n-1}
&=\sum\limits_{a=1}^{n-1}\left((\epsilon_n.k_a) \epsilon_\nu\frac{\partial}{\partial k_{a\nu}}-\frac{(\epsilon_n.k_a)^2}{k_n.k_a}k_{n\rho}\frac{\partial}{\partial k_{a\rho}}\right)\sum\limits_{b\neq l}^{n-1}\frac{k_l.k_b}{\sigma_{lb}}\notag&\\
&=\sum\limits_{b\neq l}^{n-1}\frac{1}{\sigma_{lb}}\left( 2(\epsilon_n.k_b)(\epsilon_n.k_l)
-\frac{(\epsilon_n.k_b)^2}{k_n.k_b}k_n.k_l-\frac{(\epsilon_n.k_l)^2}{k_n.k_l}k_n.k_b\right).
\end{align}
We may now compare \eqref{secondterm} with parts of the expression given by the action of $S^{(1)}$ on $M_{n-1}$ that have the same delta function support under the integral:
\begin{align}
\label{sonm}
S^{(1)}M_{n-1}
&=\bigintss \left[ \prod\limits_{c}^{n-1}{'}d\sigma_c\right](\sigma_{ij}\sigma_{jk}\sigma_{ki})\sum\limits_{l\neq i,j,k}^{n-1}\left[\prod\limits_{a\neq i,j,k,l}^{n-1}\delta\left(f_a^{n-1}\right)\right]\delta{'}\left(f_l^{n-1}\right)\left(S^{(1)}f_l^{n-1}\right)E_{n-1}\notag&\\
&+\bigintss \left[ \prod\limits_{c}^{n-1}{'}d\sigma_c\right]\left[\prod\limits_{a}^{n-1}{'}\delta\left(\sum\limits_{b\neq a}\frac{k_a.k_b}{\sigma_{ab}}\right)\right]S^{(1)}E_{n-1}.
\end{align}
Comparing \eqref{secondterm} with \eqref{sonm} we isolate the terms on the same support to be:
\begin{align}
A\equiv&\bigintss D\sigma
\sum\limits_{l\neq i,j,k,n}\left[ \prod\limits_{a\neq i,j,k,n,l} \delta\left(
f_a^{n-1}\right)
\right]
\delta{'}\left(
f_l^{n-1}
\right)
\bigintss d\sigma_n \frac{k_n.k_l}{\sigma_{ln}}\frac{\left(\sum\limits_{c\neq n}\frac{\epsilon_n.k_c}{\sigma_{nc}}\right)^2}{\left(\sum\limits_{b\neq n}\frac{k_n.k_b}{\sigma_{nb}}\right)}E_{n-1}&\\
B\equiv&\bigintss D\sigma\sum\limits_{l\neq i,j,k}^{n-1}\left[\prod\limits_{a\neq i,j,k,l}^{n-1}\delta\left(f_a^{n-1}\right)\right]\delta{'}\left(f_l^{n-1}\right),\notag\\
&\hspace{1.5in}\times\left(\sum\limits_{b\neq l}^{n-1}\frac{1}{\sigma_{lb}}\left( 2(\epsilon_n.k_b)(\epsilon_n.k_l)
-\frac{(\epsilon_n.k_b)^2}{k_n.k_b}k_n.k_l-\frac{(\epsilon_n.k_l)^2}{k_n.k_l}k_n.k_b\right)\right)E_{n-1}.\notag&\\
\end{align}
Proving that $A=B$ is equivalent to showing that:
\begin{align}
&\bigintss d\sigma_n \frac{k_n.k_l}{\sigma_{ln}}\frac{\left(\sum\limits_{c\neq n}\frac{\epsilon_n.k_c}{\sigma_{nc}}\right)^2}{\left(\sum\limits_{b\neq n}\frac{k_n.k_b}{\sigma_{nb}}\right)}\notag\\
&=\sum\limits_{d\neq n,l}\bigointss\limits_{|\sigma_n-\sigma_d|=\epsilon} d\sigma_n \frac{k_n.k_l}{\sigma_{ln}}\frac{\left(\sum\limits_{c\neq n}\frac{\epsilon_n.k_c}{\sigma_{nc}}\right)^2}{\left(\sum\limits_{b\neq n}\frac{k_n.k_b}{\sigma_{nb}}\right)}+
\bigointss\limits_{|\sigma_n-\sigma_l|=\epsilon}\sigma_n \frac{k_n.k_l}{\sigma_{ln}}\frac{\left(\sum\limits_{c\neq n}\frac{\epsilon_n.k_c}{\sigma_{nc}}\right)^2}{\left(\sum\limits_{b\neq n}\frac{k_n.k_b}{\sigma_{nb}}\right)}\notag\\
&=-\sum\limits_{c\neq l,n}\frac{1}{\sigma_{lc}}\frac{(\epsilon.k_c)^2k_n.k_l}{k_n.k_c}+\sum\limits_{c\neq n,l}\frac{2(\epsilon_n.k_l)(\epsilon_n.k_c)}{\sigma_{lc}}
-\sum\limits_{c\neq n,l}\frac{1}{\sigma_{lc}}\frac{(\epsilon_n.k_l)^2\epsilon_n.k_c}{k_n.k_l}.
\end{align}
We must now consider the first order expansion of the determinant in the last term of \eqref{exp1}. For simplicity we will choose a gauge fixing condition such that $k_n.\epsilon_a=0$. The first term in the Taylor expansion of $E_n$ is obtained by computing the derivative of the determinant  with respect to $\lambda$ evaluated at $\lambda=0$. The derivative of the determinant is obtained as a sum on $i$, of the determinant of matrices where the $i$th row is differentiated with respect to $\lambda$:
\begin{align}
\label{detexp}
\left.\frac{\partial E_n}{\partial\lambda}\right|_{\lambda=0}&=
\sum\limits_{a=1}^{n-1}\left(\frac{k_n.k_a}{\sigma_{an}}(-1)^{a+n}\tilde{\psi}_{n}^{a}+
\frac{k_n.\epsilon_a}{\sigma_{an}}(-1)^{n}\tilde{\psi}_{a+n}^{a}\right)\notag\\
&+\sum\limits_{a=n+1}^{2n-1}\left(\frac{\epsilon_{a-n}.k_n}{\sigma_{{a-n},n}}(-1)^{a+n}\tilde{\psi}_{n}^{a}-
\frac{\epsilon_{a-n}.k_n}{\sigma_{a-n,n}}(-1)^{n}\tilde{\psi}_{a-n}^{a}\right)\notag\\
&+C_{nn}(-1)^{n}\tilde{\psi}_n^{2n}\notag\\
&=
\sum\limits_{a=1}^{n-1}\left(\frac{k_n.k_a}{\sigma_{an}}(-1)^{a+n}\tilde{\psi}_{n}^{a}\right)
+C_{nn}(-1)^{n}\tilde{\psi}_n^{2n}\notag\\
&\equiv T_1+T_2,
\end{align}
where 
\begin{align}
\tilde{\psi}^a_b\equiv\left\{
\begin{array}{cc}
\frac{4\det(\Psi^{12a}_{12b})}{(\sigma_{1}-\sigma_2)^2} & \forall a,b\notin \{1,2\} \\ 
0 & a\in\{1,2\} \vee b\in\{1,2\}
\end{array} \right. .
\end{align}
and the superscript $a$ denotes removing row $a$ and the subscript $b$ denotes removing column $b$. For simplicity we have chosen i=1 and j=2 in \eqref{deteqn}.
Using the properties of the determinant we may expand each term in \eqref{detexp} as follows:
\begin{align}
T_1&=\sum\limits_{a=1}^{n-1}\frac{k_n.k_a}{\sigma_{an}}(-1)^{a+n}\tilde{\psi}_{n}^{a}\notag\\
&=-C_{nn}\sum\limits_{a=1}^{n-1}\frac{k_n.k_a}{\sigma_{an}}\left(
\sum\limits_{b=1}^{n-1}\frac{\epsilon_n.k_b}{\sigma_{nb}}(-1)^{a+b}\psi^a_b-
\sum\limits_{b=n+1}^{2n-1}\frac{\epsilon_n.\epsilon_{b-n}}{\sigma_{n,b-n}}(-1)^{a+b}\psi^a_b
\right),\\
T_2&=C_{nn}(-1)^{n}\tilde{\psi}_{n}^{2n}\notag\\
&=C_{nn}\left(
\sum\limits_{a=1}^{n-1}\frac{\epsilon_n.k_a}{\sigma_{an}}
\sum\limits_{b=1}^{n-1}\frac{k_n.k_b}{\sigma_{nb}}(-1)^{a+b}\psi^a_b
-
\sum\limits_{a=n+1}^{2n-1}\frac{\epsilon_n.\epsilon_{a-n}}{\sigma_{a-n,n}}
\sum\limits_{b=1}^{n-1}\frac{k_n.k_b}{\sigma_{nb}}(-1)^{a+b}\psi^a_b
\right),
\end{align}
where we denote removing the rows and columns $n$ and $2n$ by dropping the tilde sign.
Note that we are interested in computing the following expression:
\begin{align}
\bigintss d\sigma_n\delta \left(\sum\limits_{b\neq n}\frac{k_n.k_b}{\sigma_{nb}}\right)(T_1+T_2)
\end{align}
By shifting the summation variables and noting that $\psi^a_b=-\psi^b_a$ we may rewrite this expression as:
\begin{align}
&2\sum\limits_{a=1}^{n-1}\sum\limits_{b=1}^{n-1}(k_n.k_a)(\epsilon_n.k_b)(-1)^{a+b}\psi^{a}_{b}I_{ab}\notag\\
&-2\sum\limits_{a=1}^{n-1}\sum\limits_{b=1}^{n-1}(\epsilon_n.\epsilon_{b})(k_n.k_a)(-1)^{a+b+n}\psi^{a}_{b+n}
I_{ab}
\end{align}
where $I_{ab}$ is given by the following :
\begin{align}
\label{iab}
I_{ab}&\equiv\bigintss d\sigma_n\frac{C_{nn}}{\left(\sum\limits_{c\neq n}\frac{k_n.k_c}{\sigma_{nc}}\right)}\frac{1}{\sigma_{an}\sigma_{bn}}
\\
&=\left\{
\begin{array}{cc}
\frac{\epsilon_n.k_a}{k_n.k_a}\frac{1}{\sigma_{ab}}+\frac{\epsilon_n.k_b}{k_n.k_b}\frac{1}{\sigma_{ba}} & a\neq b\notag \\ 
\sum\limits_{d\neq n,a}\frac{1}{\sigma_{ad}}\left(
\frac{\epsilon_n.k_d}{k_n.k_a}-\frac{(k_n.k_d)(\epsilon_n.k_a)}{(k_n.k_a)^2}
\right) & a=b
\end{array} 
\right. .
\end{align}
Here we have used the fact that the integrand does not contain a pole at $\infty$ and performed a contour deformation to evaluate the integral. Using \eqref{iab} and the fact that $\psi^a_a=0$ (since it is the determinant of an anti-symmetric matrix with an odd number of rows and columns) we obtain the following expression:
\begin{align}
\label{expansione}
\bigintss d\sigma_n\delta \left(\sum\limits_{b\neq n}\frac{k_n.k_b}{\sigma_{nb}}\right)&(T_1+T_2)\notag\\
&=2\sum\limits_{a=1}^{n-1}\sum\limits_{\stackrel{b=1}{b \neq a}}^{n-1}
\frac{-1}{\sigma_{ab}}\left(\frac{\epsilon_n.k_b}{k_n.k_b}-\frac{\epsilon_n.k_a}{k_n.k_a}\right)
(k_n.k_a)(\epsilon_n.k_b)
(-1)^{a+b}\psi^{a}_{b}\hspace{2in}\notag\\
&+2\sum\limits_{a=1}^{n-1}\sum\limits_{\stackrel{b=1}{b \neq a}}^{n-1}
\frac{1}{\sigma_{ab}}
\left(\frac{\epsilon_n.k_b}{k_n.k_b}-\frac{\epsilon_n.k_a}{k_n.k_a}\right)
(\epsilon_n.\epsilon_{b})(k_n.k_a)
(-1)^{a+b+n}\psi^{a}_{b+n}\notag\\
&+2\sum\limits_{a=1}^{n-1}\sum\limits_{\stackrel{b=1}{b \neq a}}^{n-1}
\frac{-1}{\sigma_{ab}}
\left(\frac{\epsilon_n.k_b}{k_n.k_a}-\frac{(k_n.k_b)(\epsilon_n.k_a)}{(k_n.k_a)^2}\right)
(\epsilon_n.\epsilon_{a})(k_n.k_a)
(-1)^{n}\psi_{a+n}^{a} .
\end{align}
\section{The action of $S^{(1)}$ on the determinant}
\label{secs1ondet}
We now consider last term in \eqref{sonm} which contains $S^{(1)}E_{n-1}$ by considering the action of this operator on the determinant.

\begin{align}
S_b^{(1)}E_{m}&=
\sum\limits_{\stackrel{a=1}{a\neq b}}^m
\left(
\frac{S_b^{(1)}(k_b.k_a)}{\sigma_{ab}}(-1)^{a+b}\psi_b^a+
\frac{S_b^{(1)}(k_b.\epsilon_a)}{\sigma_{ab}}(-1)^{m}\psi_{a+m}^a+
\frac{S_b^{(1)}(\epsilon_b.k_a)}{\sigma_{ab}}(-1)^{a+b+m}\psi_{b+m}^a
\right)\notag\\
&+\sum\limits_{\stackrel{a=m+1}{a\neq b+m}}^{2m}
\left(
\frac{S_b^{(1)}(k_b.\epsilon_{a-m})}{\sigma_{b,a-m}}(-1)^{m}\psi^{a}_{a-m}-
\frac{S_b^{(1)}(k_b.\epsilon_{a-m})}{\sigma_{b,a-m}}(-1)^{a+b}\psi^{a}_{b}\right.\notag\\&\left.-
\frac{S_b^{(1)}(\epsilon_b.\epsilon_{a-m})}{\sigma_{b,a-m}}(-1)^{a+b+m}\psi_{b+m}^a
\right)\notag\\
&+\sum\limits_{\stackrel{a=1}{a\neq b}}^{m}
\left(
\frac{S_b^{(1)}(k_b.k_a)}{\sigma_{ba}}(-1)^{a+b}\psi^b_a+
\frac{S_b^{(1)}(\epsilon_b.k_a)}{\sigma_{ba}}(-1)^{a+b+m}\psi^{b+m}_a
\right)\notag\\
&+\sum\limits_{\stackrel{a=m+1}{a\neq b+m}}^{2m}
\left(
\frac{S_b^{(1)}(k_b.\epsilon_{a-m})}{\sigma_{b,a-m}}(-1)^{a+b}\psi^{b}_{a}+
\frac{S_b^{(1)}(\epsilon_b.\epsilon_{a-m})}{\sigma_{b,a-m}}(-1)^{a+b+m}\psi^{b+m}_a
\right)\notag\\
&+2(S_b^{(1)} C_{bb})(-1)^m \psi^{b+m}_b ,
\end{align}
Where $S_b^{(1)}=\frac{E_{\alpha\nu}k_b^\alpha k_{n,\mu} \hat{J}_b^{\mu\nu}}{k_n.k_b}$ acts on particle b as a generator of Lorentz transformations. By shifting the summation variables and using $\psi^{b}_{a}=-\psi^{a}_{b}$ one obtains the following expression:
\begin{align}
\label{sonpsi}
S^{(1)}E_{m}&=
2\sum\limits_{\stackrel{a=1}{a\neq b}}^m
\sum\limits_{b=1}^m \frac{1}{{\sigma_{ab}}}
\left(
S_b^{(1)}(k_b.k_a)(-1)^{a+b}\psi_b^a+
S_b^{(1)}(\epsilon_b.\epsilon_a)(-1)^{a+b}\psi_{b+m}^{a+m}\right.\notag\\&\left.+
\left[S_b^{(1)}(\epsilon_b.k_a)+S_a^{(1)}(k_a.\epsilon_b)\right](-1)^{a+b+m}\psi_{b+m}^a
\right.\notag\\&\left.
+\left[S_b^{(1)}(k_b.\epsilon_a)+S_a^{(1)}(\epsilon_a.k_b)\right](-1)^{m}\psi_{a+m}^a
\right) .
\end{align}

$S_b$ acts on $k_b^\beta$ with the orbital part of the generator and on $\epsilon_b^\beta$ with the spin part of the generator in the vector representation:
\begin{align}
\label{sb1}
S_b k_b^\beta&=\frac{E_{\alpha\nu}k_b^\alpha k_{n,\mu}}{k_n.k_b}k_b^{[\mu}\frac{\partial k_b^\beta}{\partial k_{b,\nu]}}\\
\label{sb2}
S_b \epsilon_b^\beta&=\frac{E_{\alpha\nu}k_b^\alpha k_{n,\mu}}{k_n.k_b}\left(\eta^{\nu\beta}\delta^\mu_\sigma-\eta^{\mu\beta}\delta^\nu_\sigma\right)\epsilon_b^\sigma
\end{align}
Taking $m=n-1$ and using \eqref{sb1} and \eqref{sb2} on \eqref{sonpsi} and the gauge fixing condition we obtain \eqref{expansione}, thus completing the proof of the conjecture for tree level graviton scattering amplitudes.
\section*{Acknowledgements}
We would like to thank Freddy Cachazo for suggesting this problem as well as insightful comments and suggestions through out various stages of this calculation. This work is supported by Natural Sciences and Engineering Research Council of Canada.

\end{document}